\documentclass{article} 
\pdfoutput=1
\usepackage{amssymb, amsmath, amsthm}
\usepackage{graphicx, subfigure}

\title{Lower Bounds for the Complexity of the Voronoi Diagram of Polygonal 
 Curves under the Discrete Fr{\'e}chet Distance}

\author{Kevin Buchin\thanks{Institute of Computer Science,
    Freie Universit{\"a}t Berlin, Germany
    {\tt \{buchin, mbuchin\}@inf.fu-berlin.de}}
    \and Maike Buchin\footnotemark[1]}

\theoremstyle{plain}
\newtheorem{thm}{Theorem}
\newtheorem{lem}{Lemma}

\newtheorem{conj}{Conjecture}

\theoremstyle{definition}

\theoremstyle{remark}

\def\eps{\varepsilon}
\def\R{\mathbb{R}}
\def\fr{Fr\'echet distance}
\def\dfr{discrete Fr\'echet distance}

\begin{document}
\maketitle

\begin{abstract}
We give lower bounds for the combinatorial complexity of the Voronoi diagram of 
polygonal curves under the discrete Fr{\'e}chet distance. 
We show that the Voronoi diagram of $n$ curves in $\R^d$ with $k$ vertices each, 
has complexity $\Omega(n^{dk})$ for dimension $d=1,2$ and $\Omega(n^{d(k-1)+2})$ 
for $d>2$.
\end{abstract}

\section{Introduction}
Important distance measures for polygonal curves are the \fr,
and its variant, the \dfr. The \fr\ can be computed in 
$O(k^2\log k)$ time for two polygonal curves with $k$ vertices~\cite{ag-cfdpc-95}
and the \dfr\ in $O(k^2)$ time~\cite{em-cdfd-94}. 

Consider the following scenario: A set $S$ of $n$ polygonal curves in $\R^d$, 
each with at most $k$ vertices, is given. The task is to find for several query 
curves the most similar curve in $S$ under the (discrete) \fr. In this setting, 
Bereg, Gavrilova, and Zhu~\cite{bgz-vddfd-07} propose to compute the Voronoi diagram of the given 
set of curves under the (discrete) \fr\ and then to locate samples 
of the transformed query curve in this. 
The Voronoi diagram of polygonal curves can be represented using 
the correspondence 
\begin{eqnarray*} 
\text{polygonal curve with $k$ vertices in }\R^d 
&\leftrightarrow& \text{point in }\R^{dk}\\
\langle(x_{11},\ldots,x_{1d}),\ldots,(x_{k1},\ldots,x_{kd})\rangle
&\leftrightarrow& (x_{11},\ldots,x_{1d},\ldots,x_{k1},\ldots,x_{kd}).
\end{eqnarray*} 

However, very little is known about the Voronoi diagram of polygonal 
curves under the (discrete) \fr. 
Recently, Bereg et al.~\cite{bgz-vddfd-07} have shown for the \dfr\  
an upper bound of $O(n^{kd+\eps})$ and a lower bound of 
$\Omega(n^{\lceil\frac{k+1}{2}\rceil})$ for the complexity of the Voronoi 
diagram of $n$ polygonal curves in $\R^d$, for $d=2,3$, 
with at most $k$ vertices each. 

We prove the following lower bounds:
\begin{thm}\label{thm:lower-bound}
 For any $d,k,n$, there is a set of $n$ polygonal curves in $\R^d$ with 
 $k$ vertices each 
 whose Voronoi diagram under the discrete Fr{\'e}chet distance has 
 combinatorial complexity $\Omega(n^{dk})$ for $d=1,2$ and $k \in \mathbb{N}$ 
and complexity $\Omega(n^{d(k-1)+2})$ for $d>2$ and $k \in \mathbb{N}$. 
\end{thm}

Our lower bounds significantly improve the lower bounds of Bereg et al.~\cite{bgz-vddfd-07}. 
For dimension $2$ the bound matches (up to $\eps$) their upper bound. 

Although Bereg et al.~\cite{bgz-vddfd-07} formulate the upper bound 
only for dimensions $2$ and $3$, their
proof generalizes to other dimensions yielding an
upper bound of $O(n^{d\cdot k + \eps})$ for $d,k \in \mathbb{N}$.
Thus, for dimensions $d=1,2$ the upper and lower bounds match (up to $\eps$), while
for $d>2$ a gap of $n^{d-2}$ between the lower and upper bound remains.
For $d>2$ our lower bound construction is a generalization of our
two-dimensional construction. One part of the generalized construction is
still inherently two-dimensional. We assume that by finding a suitable generalization
of this part or by avoiding it, the gap of $n^{d-2}$ can be closed. Therefore
we conjecture that the upper bound is tight (up to $\eps$).
   
\begin{conj} 
For any $d,k,n\in \mathbb{N}$ 
the Voronoi diagram of $n$ polygonal curves in $\R^d$ with at most $k$ 
vertices each has combinatorial complexity $\Omega(n^{dk})$ (as function in $n$).  
\end{conj}

In the following, we always use the parameters $n,d,$ and $k$ to denote 
$n$ input curves in $\R^d$, each with at most $k$ vertices.  
We give lower bounds $\Omega(f_{d,k}(n))$ on the combinatorial complexity 
of the Voronoi diagram by showing that it contains at least $f_{d,k}(n)$ 
\emph{Voronoi regions}. By a Voronoi region we mean a set of curves 
with a common set of nearest neighbors under the \dfr\ in the given set
of input curves.

\section{Lower Bounds}
We show the lower bounds in Theorem~\ref{thm:lower-bound} first for dimension 
$d=1$ (Lemma~\ref{lem:lb1d}) and then for dimensions $d\geq 2$ (Lemma~\ref{lem:lb2d}). 
For both lower bounds we construct a set $S$ of $n$ curves.
Then we construct $f(n)$ query curves which  
all lie in different Voronoi regions of the Voronoi diagram of $S$. 
This implies that the Voronoi diagram has complexity $\Omega(f(n))$. 

\begin{lem}\label{lem:lb1d}
 For all $n$ and $k$, there is a set of $n$ polygonal curves in $\R^1$ with $k$ 
 vertices each whose Voronoi diagram under the discrete Fr{\'e}chet distance has 
 at least $\lfloor \frac{n}{k} \rfloor^k$ Voronoi regions. 
\end{lem}

\begin{proof}
We construct a set $S$ of $n$ curves with $k$ vertices each
for $n = m \cdot k$ with $m\in \mathbb{N}$.
$S$ will be a union of $k$ sets $S_1,\ldots,S_k$ of $m$ curves each. 
We show that the Voronoi diagram of $S$ contains $m^k$ 
Voronoi regions.

\begin{figure}[]
  \centering
  \includegraphics[width=0.95\textwidth]{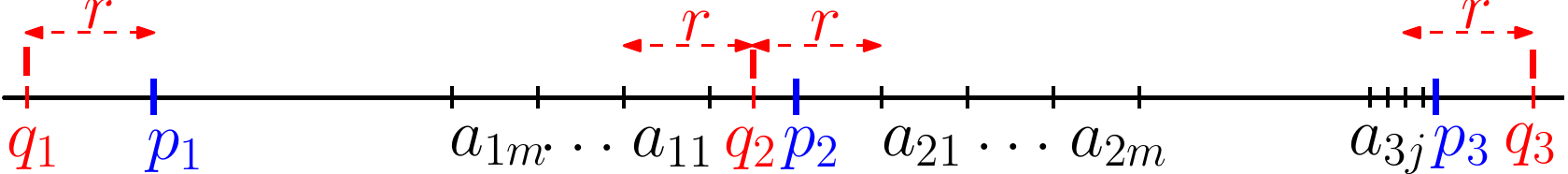}
  \caption{Construction for $d=1$ and $k=3$.}
  \label{fig:1d}
\end{figure}

The construction for $k=3$ is shown in Figure~\ref{fig:1d}.
We place $k$ points $p_1,\ldots,p_k$ with distance $2m$ between consecutive points 
on the real line. 
A curve in $S$ has the form $(p_1,\ldots,p_{i-1},p_i',p_{i+1},p_k)$ for some 
$i\in\{1,\ldots,k\}$ and point $p_i'$ close to $p_i$.
Our construction uses the following points, curves, and sets of curves.
See Figure~\ref{fig:1d} for an illustration for $k=3$.\\

\begin{tabular}{rll}
 $p_i$&$\!\!\!=\ (i-1)2m$ & for $i=1,\ldots,k$\\
 $a_{1j}$&$\!\!\!=\ p_2-j,\quad a_{2j}=p_2+j$ & for $j=1,\ldots,m$\\
 $a_{ij}$&$\!\!\!=\ p_i-j/(m+1)$ & for $i=3,\ldots,k,\ j=1,\ldots,m$\\[0.5ex]
 $S_{ij}$&$\!\!\!=\ (p_1,a_{ij},p_2,\ldots,p_k)$ & for $i=1,2,\ j=1,\ldots,m$\\
 $S_{ij}$&$\!\!\!=\ (p_1,\ldots,p_{i-1},a_{ij},p_{i+1},\ldots,p_k)$ & for $i=3,\ldots,k,\ j=1,\ldots,m$\\[0.5ex]
 $S_i$&$\!\!\!=\ \{ S_{i1},\ldots,S_{im} \}$ & for $i=1,\ldots,k$\\[2ex]
\end{tabular}
 
We claim that for all $1\leq j_1,\ldots,j_k \leq m$ a query curve $Q$ exists whose 
set of nearest neighbors in $S$ under the \dfr, denoted by $N_S(Q)$, is 
\begin{eqnarray}\label{eq:1d}
N_S(Q) = \{S_{11},\ldots,S_{1j_1},\ldots,S_{k1},\ldots,S_{kj_k}\}.
\end{eqnarray}
Since these are $m^k$ different sets, this implies 
that there are at least $m^k$ Voronoi regions.

The query curve $Q$ will have $k$ vertices $q_1,\ldots,q_k$ 
with $q_i$ close to $p_i$ for $i=1,\ldots,k$. The \dfr\ of $Q$ to any curve in $S$
will be realized by a bijection mapping each $p_i$ or $p_i'$ to $q_i$. 
Because the $p_i$ are placed at large pairwise distances, this is the best possible matching
of the vertices for the \dfr.   

Let $r=(a_{2j_2}-a_{1j_1})/2$ denote half the distance between $a_{1j_1}$ and $a_{2j_2}$. 
We choose the first vertex of $Q$ as $q_1=-r$. 
The second vertex $q_2$ we choose as midpoint between $a_{1j_1}$ and $a_{2j_2}$, i.e., 
$q_2 = (a_{1j_1}+a_{2j_2})/2$.  
Since $p_1=0$, the distance between $p_1$ and $q_1$ is $r$. 
Because all curves in $S$ start at $p_1$, this is the smallest possible \dfr\ between 
$Q$ and any curve in $S$. 
We now construct the remaining points of $Q$, such that the curves in $N_S(Q)$ 
are exactly those given in equation~\ref{eq:1d} and these have \dfr\ $r$ to~$Q$. 
   
We have already constructed $q_2$ such that it has distance at most $r$ 
to the points $a_{11},\ldots,a_{1j_1}$ and $a_{21},\ldots,a_{2j_2}$ (cf. Figure~\ref{fig:1d}). 
Now we choose the remaining points $q_i$ as $q_i=a_{ij_i}+r$ for $i=3,\ldots,k$. 
Then the point $q_i$ has distance at most $r$ to the points $p_i,a_{i1},\ldots,a_{ij_i}$
for $i=3,\ldots,k$. 
\end{proof}

\begin{lem}\label{lem:lb2d}
 For all $n,k$ and for all $d\geq 2$, there is a set of $n$ polygonal curves in $\R^d$ with $k$ vertices each 
 whose Voronoi diagram under the discrete Fr{\'e}chet distance has 
 at least $\lfloor \frac{n}{d(k-1)+2} \rfloor^{d(k-1)+2}$ Voronoi regions.
\end{lem}

\begin{proof}
We first give the construction for dimension $d=2$ and then show how to generalize it 
for $d>2$. 

\paragraph{Construction for $\mathbf{d=2}$.}
We construct the set $S$ as union of $2k=d(k-1)+2$ sets $S_1,\ldots,S_{2k}$ of $m$ curves 
each for $m\in \mathbb{N}$. 

First, we place $k$ points $p_1,\ldots,p_k$ at sufficient pairwise distance in $\R^2$, 
that is, at distance $4r$ for some distance $r>0$. 
Let $a_{11},\ldots,a_{1m}$ be points evenly distributed on the 
circle of radius $2r$ around $p_1$. 
Let $a_{21},a_{31},$ and $a_{41}$ be points evenly distributed on the circle
with radius $r$ around $p_2$. 
Let the points $a_{22},\ldots,a_{2m}$ lie on the line through $a_{21}$ and $p_2$ moved away 
from $a_{21}$ by at most $\eps>0$ as in Figure~\ref{fig:2d}. The distance $\eps$ is
sufficiently small for our construction, namely 
$\eps< r\left(\frac{1}{\cos\left(\frac{\pi}{m}\right)}-1\right)$ (assuming $m>2$).  
Place the points $a_{32},\ldots,a_{3m}$ and $a_{42},\ldots,a_{4m}$ analogously.

For $i\geq 5$ the points $a_{ij}$ are placed as follows. 
The points $a_{(i-1)j}$ and $a_{ij}$ for $i=2l$ are placed close to the point $p_l$.
Then we place $a_{(i-1)1}$ and $a_{i1}$ on the intersection of the coordinate axes originating 
in $p_l$ with the circle of radius $r$ around $p_l$. The points  $a_{(i-1)j},a_{ij}$ 
for $j\geq 2$ are placed on these axes, moved away from $a_{(i-1)1}, a_{i1}$ by at most $\delta >0$. 
The distance $\delta$ is also sufficiently small for our construction, namely  
it is $\delta\leq (\sqrt{2}-1)r$. 

\begin{figure}[]
  \centering
  \includegraphics[width=0.95\textwidth]{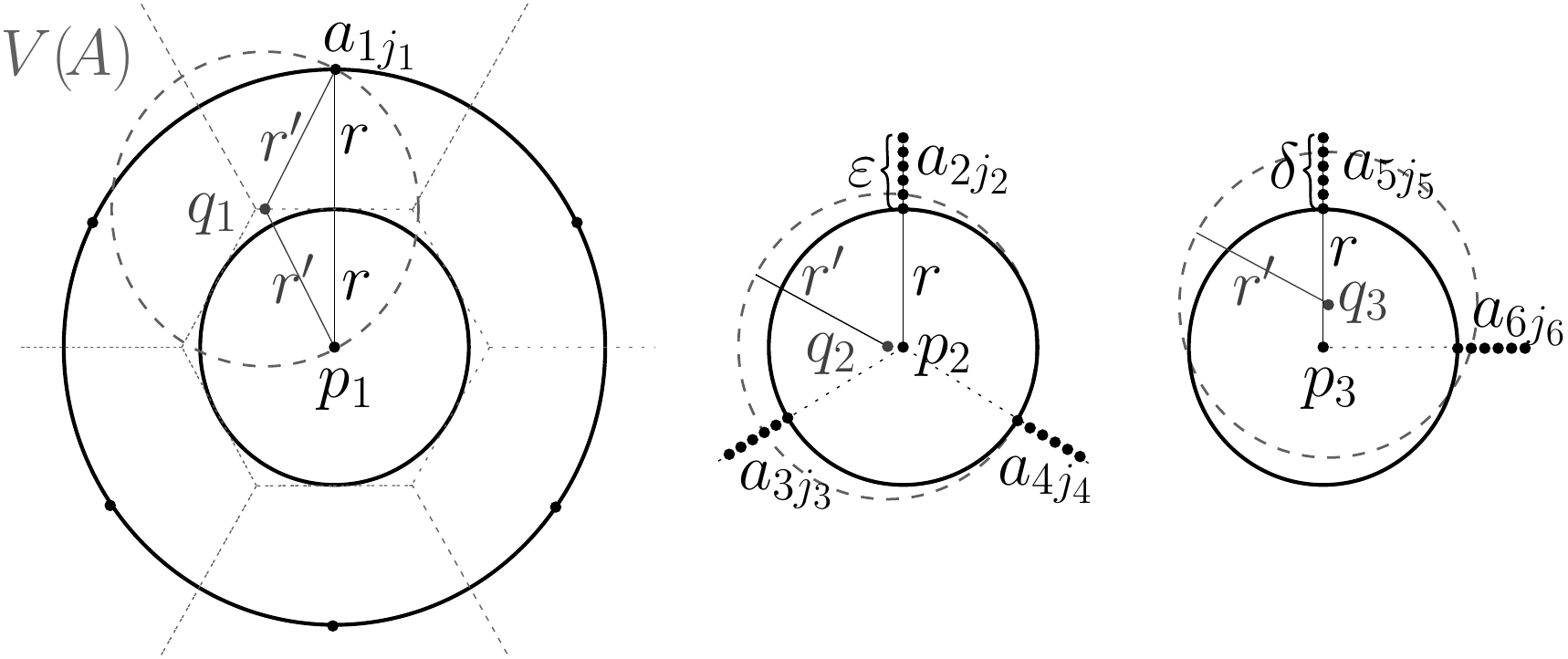}
  \caption{Construction for $d=2$ and $k=3$.}
  \label{fig:2d}
\end{figure}

We can now define the curves in $S$. As in the construction for $d=1$, 
the curves in $S$ visit all but one of the points $p_1,\ldots,p_k$, and in the 
one point deviate slightly. We define\\

\begin{tabular}{rll}
 $S_{1j}$&$\!\!\!=\ (a_{1j},p_2,\ldots,p_k)$ & for $j=1,\ldots,m$\\
 $S_{ij}$&$\!\!\!=\ (p_1,a_{ij},p_2,\ldots,p_k)$ & for $i=2,3,4,\ j=1,\ldots,m$\\
 $S_{ij}$&$\!\!\!=\ (p_1,\ldots,p_{\lceil\frac{i}{2}\rceil-1},a_{ij},p_{\lceil\frac{i}{2}\rceil},\ldots,p_k)$ & 
  for $i=5,\ldots,k,\ j=1,\ldots,m$\\[0.5ex]
 $S_i$&$\!\!\!=\ \{ S_{i1},\ldots,S_{im} \}$ & for $i=1,\ldots,2k$\\[2ex]
\end{tabular}

Again we claim that for all $1\leq j_1,\ldots,j_{2k} \leq m$ a query curve $Q$ exists 
whose set of nearest neighbors in $S$ is 
\[ N_S(Q) = \{S_{1j_1},S_{21},\ldots,S_{2j_2},\ldots,S_{(2m)1},\ldots,S_{(2m)j_{(2m)}}\}.\]
This will imply that there are at least $m^{2k}$ different 
Voronoi regions in the Voronoi diagram of $S$. 

As second point $q_2$ of $Q$ we choose the midpoint of the circle defined by the 
three points $a_{2j_2},a_{3j_3}$, and $a_{4j_4}$. Let $r'>r$ be the radius of this circle. 
Note that $r'\leq r+\eps$ and that this circle contains the point $p_2$. 
Thus, the points $p_2,a_{21},\ldots,a_{2j_2},a_{31},\ldots,a_{3j_3}$, and $a_{41},\ldots,a_{4j_4}$
have distance at most $r'$ to the point $q_2$.   

As first point $q_1$ of $Q$ we choose a point that has distance $r'$ to both 
$p_1$ and $a_{1j_1}$, and a larger distance to all other $a_{1j}$.  
Consider the Voronoi diagram of the points $p,a_{11},\ldots,a_{1m}$. 
Consider the edge between the cells of $p$ and of $a_{1j_1}$. 
Because we chose $\eps$ sufficiently small, namely
$\eps< r\left(\frac{1}{\cos\left(\frac{\pi}{m}\right)}-1\right)$, 
and because $r'\leq r+\eps$,  
there are two points in the interior of this edge with distance $r'$ to $p_1$ and $a_{1j_1}$. 
We choose $q_1$ as one of these two points. Then the distance of $q_1$ to all $a_{1j}$ 
for $j\neq j_1$ is larger than $r'$. 

Now we choose the remaining points $q_i$ of $Q$ for $i=3,\ldots,k$. 
Let $l=2i, l'=2i-1$. 
There are two circles with radius $r'$ that touch the points $a_{lj_l}$ and $a_{l'j_l'}$. 
As $q_i$ we choose the midpoint of the one circle that contains the point $p_i$. 
Then a point $a_{lj}$ or $a_{l'j}$ has 
distance at most $r'$ to $q_i$ exactly if $l\leq j_l$ or $l'\leq j_l'$, respectively. 

\paragraph{Construction for $\mathbf{d>2}$.}
The construction can be generalized to $d>2$ giving a lower bound of 
$m^{d(k-1)+2}$ for $m\cdot d(k-1)+2$ curves.

The construction at $p_1$ remains the same. 
At $p_2$ we place $d+1$ sets of points. Then one point from
each set, i.e., $d+1$ points, define a $d$-ball.
At $p_i$ for $i\geq 3$ we place $d$ sets of points. Then one
point from each set, i.e., $d$ points, define a ball of the
radius given by the choice of $q_2$.
In total, this gives us $m^{d(k-1)+2}$ choices:
$m$ choices at $p_1$, $m^{d+1}$ choices
at $p_2$ and $m^d$ choices each at $p_3,\ldots,p_k$.
\end{proof}

\section{Conclusion}
We have shown new lower bounds for the 
complexity of the Voronoi diagram of polygonal chains under the \dfr. 
Our lower bounds nearly close the gap to the known upper bounds~\cite{bgz-vddfd-07}.
However, it remains an open problem to close the gap.

We conjecture that the upper bound is tight (up to $\eps$) as we 
have shown here for dimension $d=2$. For closing the gap, consider 
how the construction for $d=2$ is generalized to $d>2$
in the proof of Lemma~\ref{lem:lb2d}. While the constructions
at the vertices $p_2,\ldots,p_k$ are replaced by higher-dimensional analogs, the construction
at $p_1$ stays the same as in the two-dimensional case. Improving
the construction at $p_1$ might close the gap.


\end{document}